\documentclass[12pt]{article}
\usepackage{amsmath,amssymb}
\newtheorem{theorem}{Theorem}
\newlength{\dinwidth}
\newlength{\dinmargin}
\newcommand{\HH}{{\mathcal{H}}}

\newcommand{\ip}[2]{\langle #1\vert#2\rangle}

\newcommand{\DD}{{\mathcal{D}}}
\newcommand{\sN}{{\mathcal{N}}}

\newcommand{\gb}{{\boldsymbol{g}}}
\newcommand{\xb}{{\boldsymbol{x}}}

\newcommand{\fA}{{\mathfrak{A}}}

\newcommand{\RR}{{\mathbb R}}

\newcommand{\CoinX}[1]{C_0^\infty({#1})}

\newcommand{\WF}{{\rm WF}\,}

\begin{document}


\title{Energy Inequalities in Quantum Field Theory\footnote{Updated and
expanded version of a contribution to the proceedings of the XIV
ICMP, Lisbon 2003.}}

\author{\Large Christopher J.\ Fewster \\[3pt]
\small Department of Mathematics,
University of York, \\[-3pt] \small Heslington,
York YO10 5DD, United Kingdom. \\[-3pt]
\small E-mail: cjf3@york.ac.uk}  
\date{\small August 2003. Revised and expanded January 2005.}

\maketitle
\begin{abstract}
Quantum fields are known to violate all the pointwise energy
conditions of classical general relativity. We review the subject
of quantum energy inequalities: lower bounds satisfied by weighted
averages of the stress-energy tensor, which may be regarded as the vestiges of the
classical energy conditions after quantisation. Contact is also made
with thermodynamics and related issues in quantum mechanics, 
where such inequalities find analogues in sharp G{\aa}rding inequalities.
\end{abstract}


\section{Introduction: Energy conditions in General \\Relativity}

In classical relativity, the energy-momentum current density
seen by an observer with four-velocity $v^b$ is defined to
be $\Pi^a=T^a_{\phantom{a}b}v^b$, where $T_{ab}$ is the stress-energy
tensor of surrounding matter.\footnote{Our metric signature is
${+}{-}{-}{-}$; units with $\hbar=c=1$ will also be adopted.}
The requirement that $\Pi^a$ should be timelike and
future-directed is known as the {\bf dominant energy condition} (DEC)
and is a natural expression of the fundamental relativistic principle that
no influence may propagate faster than light. This interpretation is
borne out by the fact that a conserved stress-energy tensor which obeys
the DEC will vanish on the domain of dependence of any closed achronal
set on which it vanishes (see Sec.~4.3 in~\cite{HawkingEllis}), so the
DEC prohibits acausal propagation of stress-energy. The DEC may,
equivalently, be formulated as the requirement that
\begin{equation}
T_{ab} u^a v^b \ge 0
\label{eq:DEC}
\end{equation}
for all timelike, future-directed $u^a$, $v^b$; it also contains (as the
special case $u^a=v^b$) the {\bf weak energy condition} (WEC), the
assertion that all timelike observers measure positive energy density. 
By continuity, this  
implies the {\bf null energy condition} (NEC), namely that
$T_{ab}k^a k^b\ge 0$ for all null $k^a$. 

The classical energy conditions are satisfied by most classical matter
models and have several important consequences.
Matter obeying the NEC tends to focus null geodesic congruences, a fact
which plays a key role in the singularity theorems~\cite{HawkingEllis},
and the WEC (respectively, DEC) is a sufficient condition for the
positivity of the ADM (respectively, Bondi) mass~\cite{Witten81,LudVic}.
However, quantum fields have long been known to violate all such
pointwise energy conditions~\cite{EGJ} and, in many models, 
the energy density is in fact unbounded from below on the class of
physically reasonable states. Moreover, the existence of negative energy
densities draws indirect experimental support from the Casimir
effect~\cite{Casimir}. In this contribution we review these
phenomena and the extent to which quantum fields satisfy weaker energy
conditions, which may be called {\bf quantum energy inequalities} (QEIs). We
also describe connections between such inequalities and 
thermodynamical stability, and some wider parallels in quantum mechanics.
Finally, the physical picture of energy condition violation which
emerges from these results is briefly discussed.

\section{The existence of negative energy densities in \\quantum field
theory}
\label{sec:enedqft}

In 1965, Epstein, Glaser and Jaffe proved that the energy density in any Wightman field
theory necessarily admits negative expectation
values (unless it is trivial)~\cite{EGJ}. Here, we give an elementary
argument for this conclusion, the basis of which goes back at least
to~\cite{FullingDavies77}, and which applies quite generally. 

Consider a theory specified by a Hilbert space $\HH$, a dense domain
$D\subset \HH$ and a distinguished vector $\Omega\in\HH$, which we call
the vacuum. In this context, a field is an operator valued distribution
on spacetime with the property that $T(f)D\subset D$ for all test
functions $f$. 
In addition, we assume only that $T$ enjoys the Reeh--Schlieder property
that no $T(f)$ can annihilate the vacuum (for nontrivial
$f$) and, for simplicity, that $T(f)$ has vanishing vacuum expectation values,
which corresponds to adopting the vacuum as the zero of energy. This is
what one would expect of the energy density in Minkowski space; one may
easily adapt the argument to cope with nonvanishing vacuum expectation
values by treating $\widetilde{T}(f)=T(f)-\ip{\Omega}{T(f)\Omega}\boldsymbol{1}$ in place
of $T(f)$. With these assumptions in place, let $f$ be any nonnegative test
function and define (for $\alpha\in\RR$)
\begin{equation}
\psi_\alpha = \cos\alpha\,\Omega + \sin\alpha
\frac{T(f)\Omega}{\|T(f)\Omega\|}\,.
\end{equation}
Then an elementary calculation yields
\begin{equation}
\ip{\psi_\alpha}{T(f)\psi_\alpha} = \zeta \sin 2\alpha + \eta (1-\cos
2\alpha)\,,
\end{equation}
where
\begin{equation}
\zeta = \|T(f)\Omega\| \qquad \hbox{and}\qquad \eta =
\frac{\ip{\Omega}{T(f)^3\Omega}}{2\|T(f)\Omega\|^2}\,.
\end{equation}
By minimising over $\alpha$, we therefore find
\begin{equation}
\inf_{\substack{\psi\in D\\ \|\psi\|=1}} \ip{\psi}{T(f)\psi} \le 
\eta-\sqrt{\eta^2+\zeta^2}  \,,
\label{eq:bd}
\end{equation}
which is negative. Of course, this argument has very little to do with
quantum field theory and almost nothing to do with energy density {\em
per se}: the key ingredient is the linear structure of
Hilbert space, and similar arguments also apply in quantum mechanics. 

We may pursue this line of reasoning a little further if we may assume that
the vacuum admits a nontrivial scaling limit for $T$ with positive
canonical dimension (see~\cite{FredHaag87} and Sec. VII.3.2
of~\cite{Haag}) and with a nontrivial two-point function in the limit.
As shown in the Appendix, one may then 
choose a sequence $f_n$ of nonnegative test functions tending
to a $\delta$-function so that $\zeta_n\to\infty$, while
$\eta_n/\zeta_n$ tends to a finite limit. It then follows from
Eq.~(\ref{eq:bd}) that the expectation value of $T$ at a point (if it
exists) is unbounded from below as the state varies in $D$.

\section{Quantum Energy Inequalities} \label{sect:qeis}

Although one cannot expect reasonable quantum
field theories to satisfy any of the pointwise classical energy
conditions, one may still hope that there
would be some vestige of these conditions in quantum field
theory: after all, they ought to emerge from the quantum field theory in the
classical limit. This leads to the conjecture that {\em smeared} energy
densities might satisfy state-independent bounds, which become
progressively weaker as the support of the smearing function shrinks,
and tighter as it grows. Bounds of this type, known as {\bf Quantum Weak
Energy Inequalities}\footnote{The original terminology was simply
``quantum inequality'' (QI); the more specific term QWEI was introduced
later~\cite{FVdirac}, as there turn out to be many other situations in
which similar bounds appear (see, e.g., Sec.~\ref{sect:qm}
and~\cite{Marecki}).} (QWEIs) were first proved by Ford~\cite{Ford91}
who was initially guided by thermodynamic considerations~\cite{Ford78}
(see also Sec.~\ref{sect:thermo}). The original bound actually concerned the energy flux,
but was soon adapted to the energy
density of the scalar and
electromagnetic fields in Minkowski space~\cite{FordRoman95,FordRoman97}. In these bounds, the
energy density is averaged along an inertial trajectory against a
Lorentzian weight; for example, the massless scalar field in
four-dimensions was shown to obey
\begin{equation}
\int dt\,\frac{\tau\langle T_{00}(t,\xb)\rangle_\psi}{\pi(t^2+\tau^2)} \ge 
-\frac{3}{32\pi^2\tau^4}
\label{eq:FRqi}
\end{equation}
for a large class of states $\psi$. The parameter $\tau$ sets the
timescale over which the average is taken; as hoped, we find that the
bound is tighter as $\tau$ increases (leading to a proof of the averaged
weak energy condition (AWEC) in the limit $\tau\to\infty$). The fact
that the bound diverges as $\tau\to 0$ is consistent with the
unboundedness from below of the energy density at a point.
Eq.~(\ref{eq:FRqi}) is of course reminiscent of the time--energy
uncertainty relation (although this is not an ingredient of the proof). 
Bounds of this type were generalised
to ultrastatic spacetimes by Pfenning and Ford~\cite{PfenningFord98},
for averages along static trajectories with the Lorentzian weight. In
curved spacetimes (or even in compact flat spacetimes) 
it is of course possible to have a constant negative
renormalised energy density, which could not satisfy a bound of
the form above. The quantity appearing in the results
of~\cite{PfenningFord98} is, instead, the difference between the renormalised energy
density in state $\psi$ and that taken in the vacuum, which we might
refer to as the normal ordered energy density. Thus these `difference' QWEIs bound the
extent to which the energy density can drop below the vacuum expectation
value. 

A different approach to QWEIs was developed by Flanagan~\cite{Flanagan97,Flan2}
for massless scalar fields in two dimensions. The resulting bound 
is not only valid for a large class of averaging weights, but is also
sharp. Yet another approach was initiated in work with
Eveson~\cite{FewsterEveson} for
averages along inertial trajectories in Minkowski space of dimension
$d\ge 2$ using a large class of weight functions. 
For example, a scalar field of mass $m\ge 0$ obeys
\begin{equation}
\int {\langle T_{00} \rangle}_{\psi} (t, \xb) \  {g(t)}^{2} dt
\ge - \frac{1}{16{\pi}^{3} }\  \int_{m}^{\infty} du \, 
|\widehat{g} (u) |^{2} u^{4} 
Q_{3}(u/m)
\label{eq:scalarQI}
\end{equation}
in four dimensions, where $Q_3:[1,\infty)\to\RR^+$ is defined by
\begin{equation} 
Q_{3}(x) = \left( 1 - \frac{1}{x^2} \right)^{1/2} \left( 1 -
\frac{1}{2x^2}  \right)  -  \frac{1}{2x^4} \ln (x + \sqrt{x^{2} -
1})
\label{eq:Q3}
\end{equation}
and obeys $0\le Q_3(x)\le 1$ with
$Q_3(x)\to 1$ as $x\to\infty$. In contrast to Flanagan's bound,
Eq.~(\ref{eq:scalarQI}) is not sharp, and differs from it by a factor of
$3/2$ in the $d=2$, $m=0$ case. Generalisations to static
spacetimes~\cite{CJFTeo}, electromagnetism~\cite{Pfenning_em} and, on a
slightly different tack, quantum optics~\cite{Marecki} are known. 

The following general QEI is based on Ref.~\cite{AGWQI} and essentially
places the argument of~\cite{FewsterEveson} in a much more general setting. 
Consider a real, minimally coupled scalar field $\Phi$ of mass $m\ge 0$
propagating on a
globally hyperbolic spacetime $(M,\gb)$. Each Hadamard state $\omega$ of the
quantum field determines a two-point function
\begin{equation}
\omega_2(x,y) = \langle\Phi(x)\Phi(y)\rangle_\omega
\end{equation}
which, in particular, satisfies the following properties:
\begin{itemize}
\item $\omega_2(\overline{F},F)\ge 0$ for all test functions $F\in\DD(M)$.
\item $\omega_2(F,G)-\omega_2(G,F)=i\Delta(F,G)$ for all
$F,G\in\DD(M)$, where $\Delta$ is the advanced-minus-retarded
fundamental solution to the Klein--Gordon equation. The important point
is that the right-hand side is state-independent.
\item The wave-front set~\cite{HormanderI} of $\omega_2$ is
constrained by $\WF(\omega_2)\subset
\sN^+\times\sN^-$, where $\sN^\pm$
is the bundle of null covectors on $M$ directed to the future ($+$) or past ($-$).
This is the {\bf microlocal spectrum condition}, which encodes the
Hadamard condition~\cite{Radzikowski96}. All Hadamard two-point
functions are equal, modulo smooth terms. 
\end{itemize}
Given a second Hadamard state $\omega^{(0)}$, which we adopt as a
reference state, the normal ordered two-point function
\begin{equation}
{:}\omega_2{:}(x,y) = \omega_2(x,y)-\omega_2^{(0)}(x,y)
\end{equation} 
is therefore smooth and symmetric and obeys
\begin{equation}
{:}\omega_2{:}(\overline{F},F)\ge
-\omega_2^{(0)}(\overline{F},F)\,.
\end{equation}
The diagonal values ${:}\omega_2{:}(x,x)$ define the Wick
square $\langle {:}\Phi^2{:}\rangle_\omega(x)$.

Now let $g$ be a smooth, real-valued function, compactly supported in a 
single coordinate patch of $(M,\gb)$, and define an averaged
Wick square by
\begin{equation}
A(g,\omega):=
\int \langle {:}\Phi^2{:}\rangle_\omega (x) g(x)^2 \,.
\end{equation}
Then, splitting the points in the definition of ${:}\Phi^2{:}$ by the
introduction of a $\delta$-function
\begin{equation}
A(g,\omega) = \int d{\rm vol}(x)\,d{\rm vol}(y)\, 
{:}\omega_2{:}(x,y) g(x)g(y)\delta_\gb(x,y)\,,
\end{equation}
where $\delta_\gb$ is the $\delta$-function on $(M,\gb)$. 
Passing to the coordinate chart containing the support of $g$, 
and writing the $\delta$-function as a Fourier integral, we find
\begin{equation}
A(g,\omega) = \int \frac{d^4k}{(2\pi)^4} \int d^4x\, d^4y\,
{:}\omega_2{:}(x,y) g(x)g(y)(\rho(x)\rho(y))^{1/2}e^{-ik\cdot(x-y)}\,,
\end{equation}
where, in these coordinates, $\rho(x)=|\det\gb_{ab}(x)|^{1/2}$.
Exploiting the symmetry of ${:}\omega_2{:}$, the $k$-integral may be
restricted to the half-space with $k_0>0$ at the expense of a factor of
$2$. We then have
\begin{eqnarray}
A(g,\omega) 
&=&  2\int_{k_0>0} \frac{d^4k}{(2\pi)^4}\,
{:}\omega_2{:}(\overline{g}_k,g_k) \nonumber \\
&\ge & -2
\int_{k_0>0}\frac{d^4k}{(2\pi)^4}\,\omega_2^{(0)}(\overline{g}_k,g_k)
\nonumber\\
&\ge & -2\int_{k_0>0}\frac{d^4k}{(2\pi)^4}\, \widehat{F}(-k,k) 
\,,
\end{eqnarray}
where $g_k(x)=e^{ik\cdot x}g(x)/\rho(x)^{1/2}$ and
$F(x,y)=g(x)g(y)(\rho(x)\rho(y))^{1/2}\omega_2^{(0)}(x,y)$. We may now
invoke the microlocal spectrum condition and Prop.~8.1.3
in~\cite{HormanderI} to show that the right-hand
side of the inequality is finite because the Fourier transform of $F$
decays rapidly in the integration region.
(We are using a nonstandard convention for the Fourier transform in which
$\widehat{f}(k)=\int dx\, f(x)e^{ik\cdot x}$.)

To convert this into a general quantum energy inequality, suppose 
$f^{ab}$ is a tensor field for which, classically,
\begin{equation}
T_{ab}f^{ab} = \sum_{j=1}^N (P_j\phi)^2\,,
\label{eq:sumofsqs}
\end{equation}
where $P_j$ are partial differential operators with smooth, real, compactly
supported coefficients. Then exactly the same argument yields a (finite) lower
bound on $\int d{\rm vol}(x) \langle {:}T_{ab}{:}\rangle_\omega(x)
f^{ab}(x)$ simply by replacing $\omega_2$ by $\sum_{j=1}^N \left(P_j\otimes
P_j \right)\omega_2$ in the definition of $F$.
Since the scalar field obeys the DEC and WEC precisely because 
the appropriately contracted stress tensor may be written in 
the `sum of squares' form~(\ref{eq:sumofsqs}), our
QEI has, as special cases, the quantum dominant/weak energy inequalities (QDEI/QWEIs). 

Several remarks are appropriate here. First, the bound depends on the
coordinate system chosen, so one has the freedom to sharpen
the bound by modifying the coordinates. Second, it is remarkable that
the bound remains finite if the support of $g$ (or $f^{ab}$) is shrunk
to a timelike curve.\footnote{Indeed, the version of this argument in~\cite{AGWQI}
considered only the case of averaging along a smooth timelike curve.} 
The same is not true for averaging along null
curves or within a spacelike slice, where one may show explicitly that the
averaged quantity is unbounded from below~\cite{FewsterRoman03,FHR}.
Third, the argument can be generalised to spin-one
fields~\cite{FewsterPfenning}. 
Fourth, restricted to static worldlines in static spacetimes, with the
reference state chosen to be a static ground state, we find 
\begin{equation}
\int dt\, \langle{:}T_{ab}u^a u^b{:}\rangle_\omega(\gamma(t)) g(t)^2 \ge -\int_0^\infty du\, Q(u)
|\widehat{g}(u)|^2\,,
\label{eq:staticQI}
\end{equation}
where $u^a$ is the four-velocity of the static worldline $\gamma$, and
$Q$ is monotone increasing and polynomially bounded.\footnote{If $\omega_0$ is
time-translationally invariant, but not a ground state, then $Q(u)$ has
a tail in the negative half-line which decays rapidly as $u\to
-\infty$.}  As a special case, we recover Eq.~(\ref{eq:scalarQI}); 
bounds of the form Eq.~(\ref{eq:staticQI}) have also appeared in
other contexts (see Sec.~\ref{sect:qm}). 
Finally, a different approach to scalar field QEIs, which also employs microlocal
techniques, can be found in~\cite{Helfer2}. 

One of the key ideas underlying the argument just given was the
positivity of the classical expression $T_{ab}f^{ab}$. The situation is
rather different in the case of a Dirac field, for which the classical
[i.e., `first quantised'] energy density is, like the Hamiltonian,
unbounded from both above and below. Positivity of the total energy
emerges for the first time after renormalisation.
For some time, this frustrated
attempts to obtain a QWEI for spin-$\frac{1}{2}$ fields. The first
success was due to Vollick~\cite{Vollick2}, who adapted Flanagan's
proof~\cite{Flanagan97} to treat massless Dirac fields in
two dimensions. Subsequently, Verch and the present author used
microlocal techniques to
establish the existence of Dirac and Majorana QWEIs in general
four-dimensional globally hyperbolic spacetimes~\cite{FVdirac}. 
However, the first explicit QWEI bound for Dirac fields in four
dimensions has only been
obtained very recently~\cite{FewsterMistry}. This bound is also of the
form~(\ref{eq:staticQI}).

So far we have only discussed free quantum fields. The situation for
general interacting fields is not yet clear (see the remarks below).
However, it is known that all unitary, positive-energy conformal fields
in two-dimensional Minkowski space obey QEIs~\cite{FewsterHollands} by
an argument based on that used by Flanagan for massless scalar fields~\cite{Flanagan97}.

Finally, we should note that there are quantum field theories which
do not satisfy QEIs. The simplest (and rather unphysical) 
example consists of an infinite number
of fields with the same mass. More serious, perhaps, is the fact that
the nonminimally coupled scalar field violates the energy conditions
even at the classical level and is not expected to obey QEIs. 
In this regard, it is worth noting that the theory of
Einstein gravity with a nonminimally coupled scalar field is 
mathematically equivalent\footnote{Equivalence holds provided the
scalar field does not take Planckian values, a regime in which the
nonminimally coupled theory is, in any case, unstable.} (in the so-called `Einstein
frame') to the theory of a {\em minimally} coupled field plus gravity (see
Ref.~\cite{FGN} for a review). In the
Einstein frame, of course, QEIs do hold. It is possible that one may
require a full theory of quantum gravity to assess the significance of the failure
of QWEIs in the usual `Jordan frame' (see~\cite{Flan04} for a careful
discussion of physics in different conformal frames).
Olum and Graham~\cite{OlumGraham} have also argued that interacting quantum fields can
violate worldline QWEIs. They consider 
two coupled scalar fields, one of which is in a domain wall configuration; away
from the wall, the second field experiences a static negative energy density
(as often occurs near mirrors). 
This suggests strongly that the existence of QEIs for worldline averages is a special feature
of the free field (or conformal fields in two dimensions~\cite{FewsterHollands}).
However, it is still plausible that QEIs exist for
spacetime averages of the stress-energy tensor. Consider a family of
smearings whose spacetime `support radius' is determined by a parameter
$\lambda$. In the situation just described, sampling over longer
timescales (say, by increasing $\lambda$) would also involve sampling
over larger spatial scales,
eventually meeting the (large) positive energy in the domain wall. 
It is certainly conceivable that the averaged energy density could still
satisfy a lower bound which tends to zero as $\lambda\to\infty$ and
diverges as $O(\lambda^{-4})$ as $\lambda\to 0^+$.

\section{Connections with thermodynamics} \label{sect:thermo}

Quantum inequalities originate from a 1978 paper of Ford 
entitled ``Quantum coherence effects and the second law of
thermodynamics''~\cite{Ford78}. Ford argued that unconstrained negative
energy fluxes (e.g., a superposition of right-moving modes with a
left-directed flux) could be used to violate the second law of
thermodynamics, by directing such a beam at a hot body to lower both its
temperature and entropy. However, 
macroscopic violations of the second
law cannot occur if the magnitude $F$ and duration $\tau$ of the negative
energy density flux are constrained by
$|F|\lesssim \tau^{-2}$ because the absorbed energy would
be less than the uncertainty of the energy of the body on the relevant
timescale. This prompted Ford to seek mechanisms within quantum field
theory which would limit negative energy fluxes and densities, and 
led ultimately to quantum inequalities of the type described in Sec.~\ref{sect:qeis}.

Recently, in work with Verch~\cite{passivity}, a new twist has been added to the connection between quantum
inequalities and thermodynamics: it turns out
that there is a rigorous converse to Ford's original
argument. We consider quantum systems in static spacetimes of
the form $\RR\times\Sigma$ where the spatial section $\Sigma$ is a compact Riemannian
manifold. The algebra of observables, $\fA$ is assumed to be a
$C^*$-algebra on which the time translations $t\mapsto t+\tau$ are assumed to
induce a strongly continuous one-parameter family of automorphisms
$\alpha_\tau$, so that $(\fA,\alpha_\tau)$ is a $C^*$-dynamical
system. We also assume that the system is endowed with an energy density
$\rho(t,\xb)$ whose spatial integral over any hypersurface
$\{t\}\times\Sigma$ generates the time evolution in the
sense that
\begin{equation}
\int_\Sigma d{\rm vol}_\Sigma(\xb) \ell([\rho(t,\xb),A]) =\left.
\frac{1}{i}\frac{d}{d\tau}\ell(\alpha_\tau(A))\right|_{\tau=0}
\end{equation}
for sufficiently large classes of observables $A\in\fA$ and continuous
linear functionals $\ell\in\fA^*$. (Precise definitions are given in~\cite{passivity}.)
One may now investigate the consequences of assuming that $\rho(t,\xb)$ satisfies
various QWEI conditions, patterned on those obeyed by quantum fields. 
In particular, a state $\omega$ of the system is said to
obey a {\bf static quantum weak energy inequality} with respect to a
class of states $\mathcal{S}$ if, for each real-valued $g\in\CoinX{\RR}$
there is a locally integrable non-negative function
$\Sigma\owns\xb\mapsto q(g;\xb)$ such that 
\begin{equation}
\int dt\, g(t)^2 \left[
\langle \rho(t,\xb)\rangle_\varphi - \langle \rho(t,\xb)\rangle_\omega\right]
\ge -q(g;\xb)
\end{equation}
for all $\varphi\in\mathcal{S}$ and $\xb\in\Sigma$. The state $\omega$
is said to be {\bf quiescent} if, in addition, each $\xb$ has an open
neighbourhood $U$ such that
\begin{equation}
\lambda\int_U d{\rm vol}_\Sigma(\xb)\, q(g_\lambda;\xb)\longrightarrow 0\qquad\hbox{as $\lambda\to
0^+$}\,, 
\end{equation}
where $g_\lambda(t) = g(\lambda t)$. (One may
regard this as a spatially averaged version of a difference AWEC.) On the
assumption that $\mathcal{S}$ is a sufficiently rich class of states, we
proved, {\em inter alia}, the following result.
\begin{theorem} If a state $\omega\in\mathcal{S}$ obeys a static QWEI
then the $C^*$-dynamical system admits a passive state. Moreover, if
$\omega$ is quiescent then it is passive.
\end{theorem}
We recall that the defining property of a {\bf passive} state of a $C^*$-dynamical
system is the impossibility of extracting net work from a system
initially in such a state by a cyclical
perturbation of the dynamics~\cite{PW}. In this sense, the passivity criterion is
an expression of the second law of thermodynamics; the force of the
above result is that thermodynamic stability may be viewed as a
consequence of QWEIs. 

The abstract results of Ref.~\cite{passivity} are complemented by a
detailed study of the free scalar field in static
spacetimes with compact spatial sections. This does not immediately fit
into our framework as the Weyl algebra describing 
the field theory is not a $C^*$-dynamical system. 
However, one may construct an auxiliary $C^*$-dynamical system
to which the structural assumptions do apply. (Such complications would
be absent for the Dirac field.)

These results lead to an interesting situation. As we have seen,
QWEIs are consequences of the microlocal spectrum condition, while
passivity is a consequence of QWEIs. Earlier work by Sahlmann and Verch~\cite{SV1}
established that states of the scalar field obeying a certain
passivity condition necessarily obey the microlocal spectrum
condition. Thus the three conditions of passivity, QWEIs and the microlocal
spectrum condition are mutually interconnected. And this is significant
because these conditions may be interpreted as a stability conditions
operating at different scales: microscopic [microlocal spectrum condition],
mesoscopic [QWEIs] and macroscopic [passivity]. 

\section{Quantum inequalities in quantum mechanics} \label{sect:qm}

A nice analogy to quantum energy inequalities may be found in the
context of Weyl quantisation. In this procedure, a classical observable
(i.e., a function on phase space) $F:\RR^{2n}\to \RR$
is represented in quantum mechanics by the operator $F_w$ on $L^2(\RR^n)$ with
action
\begin{equation}
(F_w\psi)(x)= 
\int \frac{d^ny\,d^np}{(2\pi\hbar)^n}\, F\left({\frac{x+y}{2}},p\right)
e^{ip\cdot (x-y)/\hbar}\psi(y)\,,
\end{equation}
whose expectation values may be expressed in terms of the classical
symbol $F(x,p)$ by
\begin{equation}
\langle F_w\rangle_\psi = \int \frac{d^nx\, d^np}{(2\pi)^n}\, F(x,p) W_\psi(x,p)\,,
\end{equation}
where $W_\psi(x,p)$ is the Wigner function corresponding to $\psi$:
\begin{equation}
W_\psi(x,p)=\frac{1}{\|\psi\|^2}\int d^ny\, e^{ipy}
\overline{\psi(x+\hbar y/2)}\psi(x-\hbar y/2)\,,
\end{equation}
As is well known, the Wigner function need not be everywhere positive,
so it is clear that the positivity of $F$ in no
way entails the positivity of $F_w$. This mirrors the situation with
energy density: even fields which obey the energy conditions classically
will violate them in quantum field theory. 
Given sufficient
regularity of the classical symbol $F$, however, the quantised observable $F_w$
satisfies a sharp
G{\aa}rding inequality~\cite{FePh} of the form
\begin{equation}
\langle F_w\rangle_\psi \ge -C(\hbar) \qquad\forall \psi\in\CoinX{\RR^n}\,,
\end{equation}
which, from our current standpoint, is precisely a quantum inequality. 
One may also investigate the specific example of energy densities in
quantum mechanics. As in quantum field theory, the energy density at a
point is unbounded from below, but time averages obey quantum
inequalities of a form similar to Eq.~(\ref{eq:staticQI})~\cite{EFV}.

\section{Physical Interpretation}

Quantum energy inequalities demonstrate clearly that large negative energy
densities and fluxes are associated with high frequencies (or short
length-scales, as in the Casimir effect): averaging is
required to obtain semibounded expectation values and it is crucial that
the averaging function should decay sufficiently rapidly in the
frequency domain in order that bounds of the form
Eq.~(\ref{eq:staticQI}) are finite.
Further insights have been provided by Ford and
Roman~\cite{FRqi}, who discuss positive and negative energy densities in terms
of the financial metaphor of credit and debt. Consider, for example, an
energy density taking the form
\begin{equation}
\rho(t) = A\delta(t) + A(1+\epsilon)\delta(t-T)
\end{equation}
along some inertial worldline.\footnote{This is to be regarded as a toy model for more
realistic smooth energy distributions.} Here, one can interpret $A$ as the magnitude
of `debt' incurred, $T$ as the term of the `loan' and $\epsilon$ as the
`interest rate' due on repayment. Clearly a necessary condition for
this to be the energy density of, say, a massless scalar field in two dimensions,
is that it should satisfy
\begin{equation}
\int dt\, \rho(t) g(t)^2 \ge - \frac{1}{6\pi}\int dt\, 
|g'(t)|^2\,,
\label{eq:QIqi}
\end{equation}
for all real-valued $g\in\CoinX{\RR}$,
which is Flanagan's QWEI~\cite{Flanagan97}. Constraints on $T$ and $\epsilon$
may be obtained in terms of $A$ by substituting
particular test functions $g$~\cite{FRqi}. Sharper bounds are yielded~\cite{FTqi} by
rephrasing Eq.~(\ref{eq:QIqi}) as the condition that the differential
operator
\begin{equation}
H_\rho = -\frac{d^2}{dt^2} + 6\pi \rho(t) 
\end{equation}
should be a positive quadratic form on $\CoinX{\RR}$. In the example
given, it turns out that
\begin{equation}
T< \frac{1}{6\pi A} \quad{\rm and}\quad \epsilon\ge \frac{6\pi
AT}{1-6\pi AT}\,.
\end{equation}
The two striking features are, firstly, that there is a maximum loan
term and, secondly, that the interest rate is always positive and
diverges as the maximum loan term is approached. Thus quantum fields act
so as to restore net energy density positivity locally (rather than
globally); negative energy densities are obtained only
at the expense of a nearby positive energy density of greater magnitude. 
For further results in this direction see~\cite{Pretorius,TeoWong}.

One interesting consequence of the fleeting nature of negative energy
densities is that it will be hard to observe them directly. Helfer~\cite{HelferOQI} has
argued, on the basis of various thought experiments, that quantum fields
satisfy `operational energy conditions': that is, the energy of
any measurement device capable of resolving transient negative energy
densities will necessarily be large enough that the net local energy
density will be positive. 

Finally, we mention two important applications of quantum energy
inequalities. First, they have been used to
place constraints on various ``designer spacetimes'' including warp
drive models~\cite{PF-warp} and traversable wormholes~\cite{FR-worm}
(see also~\cite{RomanReview}). 
Second, as already mentioned, Marecki has adapted quantum inequality arguments to bound
fluctuations of the electric field strength in quantum optics~\cite{Marecki}. It is
a tantalising prospect that these results may have direct relevance to
experiments in the near future.



\bigskip
{\noindent\em Acknowledgments}

\noindent Financial assistance under EPSRC grant 
GR/R25019/01 is gratefully acknowledged. 


\appendix

\section{Scaling limits}

We briefly give some more details on the statement made at the end of
Sec.~\ref{sec:enedqft}. To do this we must briefly recall the notion of a {\bf scaling limit},
introduced by Fredenhagen and Haag~\cite{FredHaag87}. Our presentation
is influenced by~\cite{SV2}. 
Consider a four-dimensional Lorentzian spacetime $(M,\boldsymbol{g})$ and fix a point $\bar{p}\in
M$ and a chart neighbourhood $\kappa:U\to\kappa(U)\subset \RR^4$ of
$\bar{p}$. We assume that $\kappa(U)$ is convex and that
$\kappa(\bar{p})=0$, and define a family of local diffeomorphisms
$\sigma_\lambda$ ($\lambda\in(0,1]$) of $U$
by
\begin{equation}
\kappa(\sigma_\lambda(p))=\lambda\kappa(p)\,.
\end{equation}
Clearly these maps form a semigroup, with the properties
$\sigma_\lambda\circ\sigma_{\lambda'}=\sigma_{\lambda\lambda'}$ and
$\sigma_1={\rm id}$, and contract $U$ to the single point $\bar{p}$ as $\lambda\to 0^+$. We also
define an action on $\DD(U^{\times n})$, i.e., test functions on the
$n$-th Cartesian power of $U$, by
\begin{equation}
(\sigma_{\lambda*} f^{(n)})(p_1,\ldots,p_n) =
f^{(n)}(\sigma_\lambda^{-1}(p_1),\ldots,\sigma_\lambda^{-1}(p_n))\,,
\end{equation}
for $(p_1,\ldots,p_n)\in \kappa(U)^{\times n}$ with $\sigma_{\lambda*}
f^{(n)}$ vanishing elsewhere. [Note that~\cite{FredHaag87} employs maps which are diffeomorphisms of
the full manifold, leading to a correspondingly more restrictive
definition of scaling limit in what follows.]

Now let $\omega^{(n)}$ be the hierarchy of $n$-point functions for the 
$T(\cdot)$ studied in Sec.~\ref{sec:enedqft} (or any other field with the
properties assumed of $T$), defined by
\begin{equation}
\omega^{(n)}(f_1\otimes\cdots \otimes f_n) = \ip{\Omega}{T(f_1)\cdots
T(f_n)\Omega}\,,
\end{equation}
and which we assume to be distributions, $\omega^{(n)}\in\DD'(M^{\times n})$.
We say that $\Omega$ has a scaling limit at $\bar{p}$ for the field $T$,
if there exists a monotone function $N:(0,1]\to[0,\infty)$ such that the limits
\begin{equation}
\widehat{\omega}^{(n)}(f^{(n)}) = \lim_{\lambda\to 0^+} N(\lambda)^n\omega^{(n)}(\sigma_{\lambda*}
f^{(n)})
\label{eq:sl}
\end{equation}
exist and are finite for all $n=1,2,3,\ldots$ and all $f^{(n)}\in
\CoinX{U^{\times n}}$, and at least one of the $\widehat{\omega}^{(n)}$
is nontrivial (i.e., not the zero distribution). 
As shown in~\cite{FredHaag87}, the $\widehat{\omega}^{(n)}$
are distributions on $U^{\times n}$ and the function
$N(\lambda)$ is `almost a power', in the sense that there exists
$\alpha$ such that 
\begin{equation}
\lim_{\lambda'\to 0} \frac{N(\lambda\lambda')}{N(\lambda')} = \lambda^\alpha
\end{equation}
for all $\lambda\in (0,1]$. The number $d=4+\alpha$ is the {\bf
canonical dimension} of $T$ at $\bar{p}$. Although our construction made use
of a particular chart, the existence of a scaling limit is
coordinate-independent, as is the function $N$. 

In what follows, we assume that the scaling limit exists at $\bar{p}$
with strictly positive canonical dimension $d$, which entails $\lim_{\lambda\to 0^+}\lambda^4 N(\lambda)=0$.\footnote{To see this, choose
$0<\lambda<1$ such that $2\lambda^d<1$, let $\lambda_0$ be such that
$N(\lambda\lambda')< 2\lambda^\alpha N(\lambda')$ for all
$0<\lambda'<\lambda_0$ and consider the sequence $\lambda_n=\lambda_0
\lambda^n$. Then it is easy to see that
\begin{equation*}
0\le \lambda_n^4N(\lambda_n)< \lambda_0^4(2\lambda^d)^n N(\lambda_0)\to 0
\end{equation*}
as $n\to\infty$. If $N$ is monotone increasing, we are done; failing
which, $N$ is monotone decreasing and we argue as follows. 
For any $\lambda'<\lambda_0$, we define $n$ by
$\lambda'\in[\lambda_{n+1},\lambda_n)$ so $n\to\infty$ as $\lambda'\to
0^+$ and then note that 
\begin{equation*}
0\le \lambda'{}^4 N(\lambda')\le \lambda_n^4 N(\lambda_{n+1}) 
\le
(2\lambda^d)^{n+1}(\lambda_0/\lambda)^4 N(\lambda_0)\to
0
\end{equation*}
as $\lambda'\to 0^+$.} We also assume that $\widehat{\omega}^{(2)}$ is nontrivial;
one may show from this that there exists non-negative $f\in\DD(U)$ such
that $\widehat{\omega}^{(2)}(f\otimes f)>0$.\footnote{Suppose
$\widehat{\omega}^{(2)}\not=0$. Then we may
find $f,g\in\DD(U)$ with $\widehat{\omega}^{(2)}(\overline{f}\otimes
g)\not=0$ because finite linear combinations of such tensor products are
dense in $\DD(U\times U)$. A polarisation argument, using the fact that $\widehat{\omega}^{(2)}$ is
manifestly positive type, enables us to find $f$ such that
$\widehat{\omega}^{(2)}(\overline{f}\otimes f)>0$, and we may take $f$
real-valued without loss [by taking real and imaginary parts and
applying Cauchy--Schwarz]. We split $f$ into positive and negative
parts, mollify to regain smoothness and apply the Cauchy--Schwarz
argument again to obtain the required statement. Continuity
of $\widehat{\omega}^{(2)}$ in the test functions is also used.} Setting 
$f_\lambda=\lambda^{-4}\sigma_{\lambda*}f$, and considering the limit
$\lambda\to 0^+$, we note that the support of
$f_\lambda$ tends to $\{\bar{p}\}$ while $\int f_\lambda \,d{\rm
vol}_{\boldsymbol{g}}$ tends to a constant, which we may normalise to
unity. Thus $f_\lambda\to \delta_{\bar{p}}$ as $\lambda\to 0^+$.

To complete the argument, we define $\zeta_\lambda=\|T(f_\lambda)\Omega\|$ and $\eta_\lambda =
\ip{\Omega}{T(f)^3\Omega}/(2\zeta_\lambda^2)$. Clearly
\begin{equation}
\lambda^8 N(\lambda)^2\zeta_\lambda^2\to \lim_{\lambda\to 0^+}
N(\lambda)^2\omega^{(2)}(\sigma_{\lambda*}f^{\otimes 2})\not=0
\end{equation}
as $\lambda\to 0^+$; hence, because $\lambda^4 N(\lambda)\to 0$, we have
$\zeta_\lambda\to\infty$. On the other hand,
\begin{equation}
\frac{\eta_\lambda}{\zeta_\lambda}
= \frac{N(\lambda)^3\omega^{(3)}(\sigma_{\lambda*}f^{\otimes
3})}{(N(\lambda)^2\omega^{(2)}(\sigma_{\lambda*}f^{\otimes 2}))^{3/2}}
\end{equation}
and therefore tends to a finite (possibly zero) limit. Comparing with
the discussion in Sec.~\ref{sec:enedqft}, we see that
\begin{equation}
\lim_{\lambda\to 0^+} \inf_{\substack{\psi\in D \\ \|\psi\|=1}}\ip{\psi}{T(f_\lambda)\psi} =
-\infty
\end{equation}
while $f_\lambda\to \delta_{\bar{p}}$. Thus the energy density at $\bar{p}$
is unbounded from below, as claimed. 

Finally, we remark that this result continues to hold even if the vacuum
expectation value $\ip{\Omega}{T(\cdot)\Omega}$ is nonvanishing, provided it
is continuous at $\bar{p}$ and
$\widetilde{T}(f)=T(f)-\ip{\Omega}{T(f)\Omega}\boldsymbol{1}$ has a
scaling limit of the required type, because the overall energy density
is merely shifted by the finite constant
$\ip{\Omega}{T(\bar{p})\Omega}$. 







\end{document}